# AN ADAPTIVE AND INTELLIGENT TUTOR BY EXPERT SYSTEMS FOR MOBILE DEVICES

Hossein Movafegh Ghadirli[1] and Maryam Rastgarpour[2]

[1]Graduate student in Computer Engineering, Young Researchers Club, Islamshahr Branch, Islamic Azad University, Islamshahr, Iran
`hossein.movafegh@iau-saveh.ac.ir`
[2]Faculty of Computer Engineering, Department of Computer, Science and Research Branch, Islamic Azad University, Saveh, Iran
`m.rastgarpour@iau-saveh.ac.ir,m.rastgarpour@gmail.com`

## ABSTRACT

*Mobile Learning (M-Learning) is an emerging discipline in the area of education and educational technology. So researchers are trying to optimize and expanding its application in the field of education. The aim of this paper is to investigate the role of mobile devices and expert systems in disseminating and supporting the knowledge gained by intelligent tutors and to propose a system based on integration of intelligent M-Learning with expert systems. It acts as an intelligent tutor which can perform three processes - pre-test, learning concept and post-test - according to characteristic of the learner. The proposed system can improves the education efficiency highly as well as decreases costs. As a result, every time and everywhere (ETEW) simple and cheap learning would be provided via SMS, MMS and so on in this system. The global intention of M-Learning is to make learning "a way of being".*

## 1. INTRODUCTION

Nowadays, researchers try to develop a new paradigm of e-learning to mobile space(M-Learning). The term "Mobile Learning" means educational application of a variety of mobile technologies. M-Learning is so regarded after advent of computer-based learning. But its implementation is more complex than computer-based learning [1]. Mobile devices and wireless networks can provide texts, video on demand, and information services. Learners would require such technology as appropriate mobile devices, and high-bandwidth wireless networks. Connectivity is also important because disconnection caused by handovers seriously affect the quality of information services.

Diversity of learners leads to decrease the efficiency of this style. In fact, some lessons must be repeated several times for a number of learners while the same courses may be removed for others [2].

Later, the researchers in pedagogy sciences concluded that the learning must be dynamic and intelligent. The fact is that an expert tutor can adapts lessons sequence and training speed with aptitude and characteristics of learner. He can also adjust the expression style with learner's mood as well as cancels the class due to incorporate mental conditions of the learner.

The most of recent systems run on desktop computers and are not designed for mobile application such as mobile phones, smart phones, PDAs, etc. On the other hand, M-Learning systems aim to improve the quality of learning by providing mobile learners with an easy, contextualized and ubiquitous access to knowledge [1].

M-Learning has some problems, like technical nature, which is related to porting existing systems to a new environment. It introduces some complex implementation issues. Namely, the wireless environment (GSM, UMTS, WLAN) is much different from the standard "wireline" one; bandwidth, delay, error rate, interference and the like, may change dramatically as the learner changes her/his location [1].

This paper presents an intelligent system to apply the abilities of expert systems. So E-learning would be efficient, adaptive and performed by mobile devices. Adaptation is very important, because the contexts would be used by millions variant learners. So the concept, which is developed for one user, isn't applicable for others [3].

The proposed system determines the learning style via a test. Then the learning process starts. Gradually, some characteristics of learner may be change by learner's progress. These improvements would be saved by system in learning process. So learner model gets closer step by step. The system can receives scientific and mental feedback of learner expertise and then change the learning style during the process. M-Learning has surpassed of Web-based learning because, it is estimated the total number of mobile phone users worldwide is over 300 million, double the number of Internet users [4]. M-Learning content is installed and supported in one place while millions learners can use it just via a mobile device. The aim of proposed system is to offer the content which the user is not aware of it.

The rest of this paper is organized as follows. Section 2 defines an intelligent tutoring system and presents some available samples. Then it deliberates M-Learning and some learning styles in section 3. Section 4 describes the proposed M-Learning system which is intelligent, adaptive and customizable. Finally this paper concludes in section 5.

## 2. INTELLIGENT TUTORING SYSTEM (ITS)

ITSs are computer-based instructional systems with educational content models. They specify what to teach, and also teaching strategies that specify how to teach [5]. In late 1960, the ITSs have moved out of academic labs and have been applied in classrooms and workplaces. Some of them have shown to be high effective [6]. Unfortunately, intelligent tutors are difficult and expensive to build whereas they are more common and have proven that to be more effective.

Intelligent systems can recognize the learner type, choose appropriate course content from knowledge base and present it to learners in proper style [2]. It also attempts to simulate a human tutor expertly and intelligently. Students using these systems usually solve problems and related sub-problems within a goal space, and receive feedback when their behaviour diverges from that learner model.

Some design factors of ITS include of component expert simulator, tutor software, learner model, modeller, and knowledge base. Fig. 1 illustrates them.

ITSs allow "mixed-initiative" tutorial interactions, where learners can ask questions and have more control over their learning.

A mobile ITS integrates three fields of research that contribute to the design and implementation of the mobile intelligent tutoring system, namely the Intelligent Tutoring Systems, Mobile Human Computer Interaction (HCI), and Mobile Learning disciplines, as shown in Fig. 2.[7]

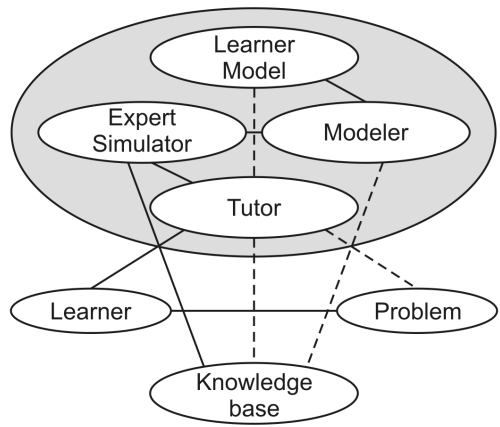

Figure 1. Intelligent Tutoring System Components [2]

Some available intelligent tutoring systems are introduced in the following.

**xTEx-Sys [8]** is actually not an intelligent tutoring system itself; it is an authoring shell supporting the development of ITSs.

**Älykkö [9]** works as a basis for M-Learning and a tutoring dialogue between students and teachers. In addition, it enables the creation of the structure and framework for M-Learning process taking place in authentic environments. Älykkö is a PHP- and MySQL-based application with automatic and semiautomatic tutoring features.

**C-POLMILE [10]** is an intelligent learning environment with an open learner model for C programming. There are two versions of the system - the desktop PC version, and a version for use on a handheld computer. The student can use whichever is most convenient at the time, and can take part in the kind of interaction which is most appropriate for their current location, the device being used, the time they have available, the likelihood of distraction in that location, etc.

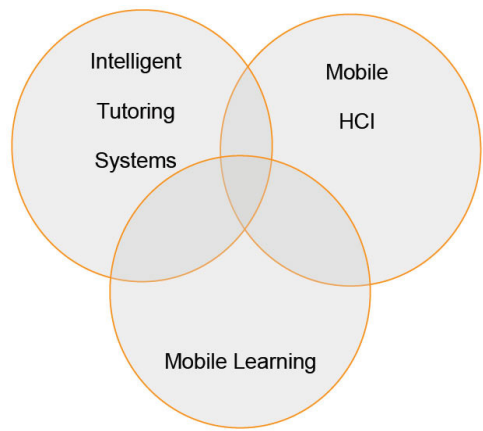

Figure 2. Integration of disciplines for the Mobile Intelligent Tutoring System [7]

The next section describes the properties of mobile devices and It also express how to learn by such devices.

23

## 3. MOBILE LEARNING

Mobile learning is an emerging discipline in the area of education. A review of M-Learning projects funded by the European Union since 2001 [11]. It confirms that mobile phones are the most frequently used device in these projects, followed by PDAs and other handhelds, with personal listening devices (e.g. iPods) receiving a little less attention. M-Learning is the point at which mobile computing and electronic learning intersect to produce an anytime, anywhere learning experience.

The field of M-Learning is approximately a decade old and has grown rapidly. As personal computer integration predicated computer-enhanced learning, a primary factor in the growth of M-Learning is the increasingly ubiquitous integration of cell phones into society [7].

The term "Mobile Devices" includes standard cell phones (those are without an operating system performing basic cellular voice communications), smart phones (those utilize an operating system providing voice services as well as additional data processing applications), and personal digital assistants (PDAs provide data processing without voice capabilities). User interaction in laptop computers is similar to desktop computers. Laptop computers are portable while desktop computers are not. These reasons lead to increase laptop users.

This system for M-Learning defines several functional levels, simplifying the design and development. So that different parties (e.g. vendors, providers, designers) can address individual levels [12]. This system enjoys all entities. So a single entity is not forced to do all tasks for developing an M-Learning system in this way.

The system is divided into four levels. In the M-Learning applications level, many new applications gets possible, and many existing electronic learning applications can be modified for a mobile environment. In the mobile user infrastructure level, the design of new M-Learning applications should consider the capabilities of the user mobile devices. In the mobile protocol level, the aim is to hide the underlying network's details from applications while providing a uniform and easy-to-use interface. In the mobile network infrastructure level, service quality primarily depends on network resources and capabilities.

### 3.1. Learning Styles

It's worth to be noted that the concept of "learning style" has emerged as a focal point of much psychological research. In our proposed learner model, we have adopted the Jackson's learning styles profiler in order to model learning styles of the learners. Jackson proposed five learning styles which are summarized in Table 1. Reader can find more details in [2].

Table 1. Summarization of the available Learning Styles [13]

| Learning style | Comment |
| --- | --- |
| Sensation Seeking (SS) | Believing that the experiences create learning. |
| Goal Oriented Achievers (GOA) | Self-confident to achieve difficult and certain target. |
| Emotionally Intelligent Achievers (EIA) | Rational and goal-oriented. |
| Conscientious Achievers (CA) | Responsible and insight creator. |
| Deep Learning Achievers (DLA) | Interested in learning highly. |

## 4. PROPOSED SYSTEM

The adaptive and intelligent tutor by expert systems is to promote the reading ability of learners based on the individual learning styles. It can estimate the reading abilities of individual learners. Then it recommends appropriate content to individual learners according to learner's simple feedback responses and difficulty parameters of the contents. It can determine learner type (especially in terms of learning style), learning content adaptively. So that it can be updated automatically with learner's characteristics and behaviour in server side. Learning all courses can be customized well at home via SMS, MMS and so on in this system. So learners can solve some examples and proper exercises ETEW. Finally he can attend in course examination which would be virtual or physical.

### 4.1. The Interface

Intelligent tutors, such as Andes, Algebra I, and AutoTutor, provide users interfaces divided into regions. Andes, an intelligent tutor for Newtonian Physics, have an interface divided into three regions/panes [14]. Similarly, Algebra I and AutoTutor developers employed a multi-region interface design [15, 16]. Users are given the specific problem or scenario which they are currently working on, a workspace, hint buttons, and other areas such as a glossary, or reference material necessary. As students use the tutors on large desktop displays, they are easily able to see all of these regions without having to manually navigate to each individually. The problem of tutor implementation in the smaller device is that the displays are small as well.

The mobile users would like to access some information without keyboard and typing the words by small fonts in a mobile tutor interface.

When the user is interacting with the mobile tutor, he should solve the problems instead of getting help. So user time is wasted. It seems that there is some interface challenges such as the limitation in user input in addition of the problems in small displays [17]. However, we focus the mobile interface.

### 4.2. Training Method

To find the role of a mobile tutor in learning, we will examine how mobile tutors can be integrated into classes and the education. The primary difference between mobile tutor and desktop tutor is the frequency and duration of applying. Now, two questions should be considered. The first question is "Is it practical to use a small size screen and to limit interaction modes?" . Another one is whether the M-Learning is as efficient as the desktop counterparts in terms of frequency and duration. However, considering that the students are often from the 5th and 6th grades, they comfortable play video games through small handheld consoles and we expect that they would be equally comfortable when using a mobile tutor [18].

Expert tutor's knowledge includes of two parts, *course knowledge* (learning content) and *learning technique*. Course knowledge is theoretic information, technical content and probably experiments which expert tutor notes. Learning technique is some experiences which he have got during teaching years [19]. An expert tutor determines learner level in according to IQ, understanding, behaviours, talent and individual characteristic like physical class. Learner level consists of "*weak*", "*slow learner*", "*smart*", "*genius*" and so on. Tutor teaches educational content corresponding to learner level in proper method such as *film*, dynamic view, and *game*. So learner level may be changed while he get feedback from learner during training. The tutor helps learner to learn by "*the best way*" in proposed system.

The expert tutor offers an education method based on learner's type. Tutor often determine different scores for variant sections according to education method. Moreover, he marks highest score to the most important section of course in all education methods.

The smallest part of any topic which can't divide more is called "concept". It is usually equivalent a lesson in physical class. Educational concepts transfer to knowledge base in this system. Then the system can distinguish all concepts and relocate all parts. Sometimes, a lesson is needed to repeat, relocate or even remove for a learner. Most of available systems guide a

25

learner to a special aim intelligently in learning process. While only a few intelligent systems provides selecting subsections of a concept for a learner.

This system uses a three layered structure to offer and implement a concept:

1. Pre-test
2. Learning concept
3. Post-test

The pre-test includes of some questions planned by an expert tutor to determine learner's primary knowledge level. The learning concept depends on learner level. So the best method to train a learner is determined in this way. Then learning process starts up. After learning is done, a post-test evaluate the learner by some questions. Figure 3 shows architecture of proposed system.

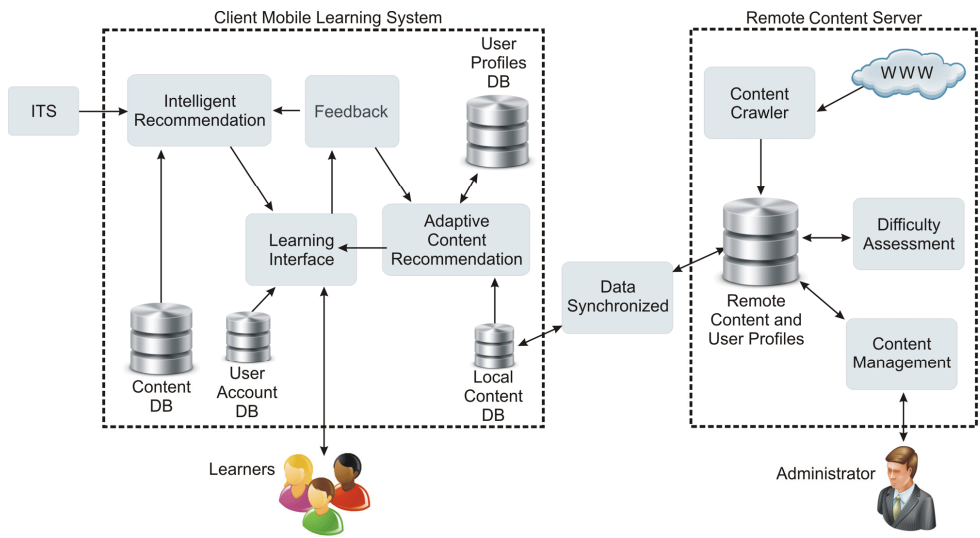

Figure 3. The Architecture of Proposed System

## 4.3. Learning Evaluation

Learner evaluation is significant. It has two levels, conceptual and objective. Evaluating in concept level refers to learner understanding of lesson concept and evaluating in objective level denotes to learner understanding of lesson topic [2]. Knowledge level of learner is determined with concept level and objective level. The tutor can extract proper questions from question base through an expert system, pre-test and post-test. He notes that a specific score is given to each question.

Selecting question should satisfy some rules. Firstly, the questions should not be repetitious even if a learner would be trained one concept several times. Secondly, the question must be planned for all sections of a concept entirely and uniformly. Thirdly, expert tutor plans questions in all level. Sequence, number and level of questions are determined according to learner level and learning type intelligently. Sum of scores is calculated and learner level is determined after answering the questions.

Table 2 presents five categories of learner's knowledge level about a concept [20]:

Table 2. Categories of Knowledge Level

| Knowledge Level | Score |
|---|---|
| Excellent | 86-100 |
| Very good | 71-85 |
| Good | 51-70 |
| Average | 31-50 |

This system updates the learner's model during progress of question answering. This system can also save last academic status of learner and all his learning records.

## 5. CONCLUSIONS

An adaptive and intelligent tutor by expert system for mobile devices was presented in this research. The focus of this paper is to outline the process involved in studying the role of mobile devices in disseminating and supporting the knowledge gained by intelligent tutors. The proposed system can adapt with learning styles (i.e. Sensation Seeking, Goal Oriented achievers, Emotionally Intelligent Achievers and Conscientious Achievers), aptitude, characteristics and behaviours of a learner. It acts as an intelligent tutor which can perform three processes - *pre-test, learning concept* and *post-test* - according to characteristic learner. This system uses expert simulator and its knowledge base at server side. It is also based on SMS, MMS and so on. It leads to be simple learning, low-cost, available everywhere and every time. Consequently thousands of students can learn simultaneous and integrated efficiently.

Despite of facilitating all schools with Internet and computer access to students, a deeper examination reveals that the presence of technology does not guarantee that it is applied efficiently. Furthermore, it solves the drawback of previous system and human expert tutor. It can improve efficiency of pedagogy and education too. In other words, it helps learners to study in "*the best way*".

**Authors**

**Hossein Movafegh Ghadirli** received his B.S. in Computer Engineering from Saveh branch, Islamic Azad University (IAU), Saveh, Iran in 2009 and He is currently a graduate student in Computer Engineering at Science and Research branch, IAU, Saveh, Iran. His overriding interest has been bringing E-Learning, M-Learning and Intelligent Tutoring Systems to improve their productivity for both government and commercial organizations. He is a member of Young Researchers Club, Islamshahr Branch, Islamic Azad University, Islamshahr, Iran. 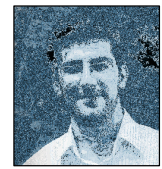

**Maryam Rastgarpour** received her B.S. in Computer Engineering from Kharazmi University, Tehran, Iran in 2003, and the M.S. in Computer Engineering from Science and Research branch, Islamic Azad University (IAU) , Tehran, Iran in 2007.She is currently a Ph.D. candidate in AI there. She is also a lecturer at Computer Department, Faculty of Engineering, Saveh branch, IAU for graduate and undergraduate students. Her research interests include in the areas of Machine Learning, Pattern Recognition, Expert Systems, E-Learning, Machine Vision, specifically in image segmentation and Intelligent Tutor System. 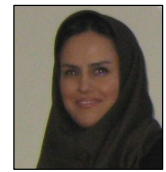